\documentclass [a4paper,12pt]{article}
\newcommand{\euro}{E}
\usepackage{multirow}
\usepackage{amsmath}
\usepackage{amssymb}
\usepackage{graphics}
\renewcommand{\thetable}{\Roman{table}}

\usepackage{sidecap}
\usepackage[sf]{caption}

\renewcommand{\figurename}{Fig.}
\renewcommand{\tablename}{Table}
\renewcommand{\thetable}{\Roman{table}}
\renewcommand{\thefigure}{\arabic{figure}}
\makeatletter
\renewcommand{\fnum@figure}{\sffamily\textbf{\figurename~\thefigure}}
\renewcommand{\fnum@table}
{\sffamily\textbf{\tablename~\thetable}}

\begin{document}

\title{Mathematical Analysis of Money in the Scope of Austerity}

\author{Peter Stallinga\\
University of The Algarve\\
e-mail: pjotr@ualg.pt}
\maketitle

\normalsize

\begin{abstract}

This summarizes the study of the financial and economic crisis in Europe. The starting questions were\\
1) Why do we have a crisis? \textit{Unde venis?}\\
2) What will be the outcome? \textit{Quo vadis?}

Here is the reasoning which touches many areas, ranging from financial to politics and from psychology and economy.
\end{abstract}

\newpage

\section{Political}

The political idea and the political measures, 'austerity', are based on a simple and single myth. Namely that economic growth is stopped when the national debt reaches a critical limit of 90\% of the GDP. This comes from a single scientific article published by two economists, Reinhart \& Rogoff. In the meantime this article has been debunked as being scientific crap [see Krugman]. Very few economists agree with the ideas. Nearly no economists think that Austerity is the answer to the crisis.\\
In the meantime -- you cannot stop a running train -- austerity will continue. Austerity is \emph{ruinous} for our society, as will be shown here. It is unnecessary and ruinous.

\section{Economy pseudo science}

The article that proved that a 90\% debt is fatal is wrong for three reasons:
\begin{enumerate}
\item They inverted cause-and effect. While (marginally) proving that "slow economic growth" and "a debt of more than 90\%" are related, the cause and effect were inverted. In fact, a debt of more than 90\% is \emph{caused} by a slow growth and not the other way around. Ergo, \emph{reducing the debt will not cause growth!} (In fact, the opposite happens, as most economists predict and what everybody can see).
\item They threw away data that were not agreeing with their thesis.
\item They analyzed the remaining data wrongly (Excel programming error).
\end{enumerate}
An example is Portugal, the country where I live. When it reached the critical threshold of 90\% debt, under pressure of the international community (read: European Union), it was forced to introduce Austerity. The result was an economy that came to a standstill and the debt skyrockets ever since then, standing at 123\% at the moment of writing this.

\section{Psychology}

The reason why politicians think the way they do is best explained by psychology. Because they themselves have little to no knowledge of economy, they use 'common sense', which is a semi-religious attitude (in the sense that you \emph{believe} things). Combined with the fact that all people try to make things simple in their head, they start analyzing the things in a general way, treating micro-economy ('house-hold' economy) and macro-economy (state economy) as the same thing. These things, however, are fundamentally different. Where a statement as "you cannot spend more than you get in, in the long run", i.e., you cannot keep on borrowing money, is valid for a family household, this is not at all true for a state economy. In fact, borrowing money is the engine of economy. Without keeping to borrow money, the economy will crash.

\section{Sociology}

Because the politicians are convinced of their own right, they select the scientific publications that agree with their ideas, while ignoring others. Then, as self-acclaimed heroes of our society, they hire the new scientists that will prove their ideas. (This is a phenomenon equal to the one used for the subject of Climate Change). After a while they have a set of papers in their hand that prove what they always 'knew'.

Yet, the ideas are incorrect. To show this, we have to study how banking works -- especially fractional-reserve banking (FRB) -- take a look at the phenomenon of interest rates, economy (specially as explained by Karl Marx and Thomas Malthus), and sociology (namely liberalism). Let's get rolling.

\section{Fractional-reserve banking}

FRB is a technique where a bank can lend more money than it has itself available ('deposited' by clients). Normally, a ratio is 9:1 is used, money lent vs. the base product of banking. 

This base product used to be gold. So, a bank could issue 9 times more 'bank notes' ('rights to gold') than it had gold in its vault. Imagine, a person comes with a sack of 1 kilo of gold. This person gets a note from the bank saying "you have deposited 1 kilo of gold in my bank. This note can be exchanged for that 1 kilo of gold any time you want". But it can legally give this same note to 8 more people! 9 notes that promise 1 kilo of gold for every kilo of gold deposited. Banks are masters of promising things they in no way whatsoever can ever fulfill. And, everybody knows it. And, still we trust the banks. It is an amazing mass denial effect. We trust it, because it gives us wealth. This confidence in the system is what is, actually, \emph{essential} in the economy. Our civilization depends on the low-morality of the system and our unwavering confidence in it. You are allowed to lie even if the lie is totally and utterly obvious and undeniably without a shred of doubt a lie.

In modern times, the gold standard has been abandoned, because it limits the game. Countries with the most advanced financial structures are the richest. Abandoning the gold standard creates enormous wealth. Rich, advanced nations, therefore, have abandoned the gold standard. In modern banks, no longer gold, but money itself is the base. That is, the promissory notes promise … promissory notes. It is completely air. Yet, it works, because everybody trusts it'll work.

Moreover, banks no longer issue bank notes themselves, except the central bank. The 'real' money of the central bank is called 'base money' (M0 or 'Tier 1') and serves as 'gold' in modern banks. The 'bank notes' from the bank promise bank notes from the central bank.

Banks use this base money no longer to directly print money (bank notes), but something that is equivalent, namely to lend money to their clients by just adding a number on their account. This, once again, works because everybody trusts it works. But is has become even thinner than air. It is equal to vacuum. There is no physical difference whatsoever anymore between having money and not having it. If I have 0 on my account, or 10000000000000000 dollars, I have the same size information on the computer of my bank. The same number of bytes (however many they may be). I just hope that one day a tiny random fluctuation occurs in their computer and sets me the first bit to a '1' (unless it is the 'sign' bit, of course). Nobody would notice, since there is nowhere money disappearing in the world. Simply more vacuum has been created.

But, it gets even worse. This newly created 'money' (the number on an account of bank A) can be deposited in other banks (write a check, deposit it, or make a bank transfer to bank B). In this other bank B, it can again be used as a base for creating money by adding a number to peoples bank account. As long as a certain amount of base money (M0, or 'Tier 1') is maintained.\\
As a side mark, note that bankers do not understand the commotion of the people in calling their rewards astronomical, since they know -- in contrast to the people that think that money represents earning based on hard work -- that money is vacuum. Giving a bonus to the manager in the form of adding a couple of zeros to her account in her own bank is nothing but air. The most flagrant case of self-referential emptiness is the bank that was bought with its own money.

In this way, the money circulating in the economy can be much larger than the base money (of the central bank). And, all this money is completely air. The amount of money in the world is utterly baseless. Since it is air, moreover an air-system that is invented to facilitate the creation of wealth, we can intervene in the system in \emph{any} way we want, if we see that this intervention is needed to optimize the creation of wealth. Think of it like this: the money and the money system was invented to enable our trade to take place. If we see that money no longer serves us (but we, instead, seem to serve the money) and decide to organize this trade in another way, we can do so without remorse. If we want to confiscate money and redistribute it, this is morally justified if that is what it takes to enable the creation of wealth.\\
Especially since, as will be shown, there is no justice in the distribution. It is not as if we were going to take away hard-earned money from someone. The money is just accumulated on a big pile. Intervention is adequate, required and justified. Not intervening makes things much worse for everybody.

Important to make this observation: All money thus circulating in the world is borrowed money. Money is nothing less and nothing more than debt. Without lending and borrowing, there is no debt and there is no money. Without money, there is no trade and no economy. Without debt, the economy collapses. The more debt, the bigger the economy. If everybody were to pay back his/her debt, the system would crash.

Anyway, it is technically not possible to pay back the money borrowed. Why? Because of the interest rates.

\section{Interest rates}

Interest is the phenomenon that somebody who lends money -- or actually whatever other thing -- to somebody that borrows it, wants more money back than it gave. This is \emph{impossible}.

To give you an example. Imagine we have a library, and this library is the only entity in the world that can print books. Imagine it lends books to its customers and after one week, for every book that it lent out, it wants two back. For some customers it may still be possible. I may have somehow got the book from my neighbor (traded it for a DVD movie?), and I can give the two books the library demands for my one book borrowed. But that would just be passing the buck around; now my neighbor has to give back to the library two books, where he has none.\\
This is how our economy works. And, to explain you what the current solution is of our society is that the library says "You don't have two books? Don't worry. We make it a new loan. Two books now. Next week you can give us four".

This is the system we have. Printing money ('books') is limited to banks ('libraries'). The rest borrow the money and in no way whatsoever -- absolutely out of the question, fat chance, don't even think about it -- is it possible to give back the money borrowed \emph{plus} the interest, because this extra money simply does not exist, nor can it be created by the borrowers, because that is reserved to the lenders only. Bankrupt, \emph{unless} these lenders refinance our loans by new loans.

When explaining this to people, they nearly always fervently oppose to this idea, because they think that with money new wealth can be created, and thus the loan can be paid back including the interest, namely with the newly created wealth. This, however, is wrong thinking, because wealth and the commodity used in the loan are different things.\\
Imagine it like this: Imagine I lend society 100 euros from my bank with 3\% interest. The only euros in circulation, since I am the only bank. Society invests it in tools for mining with which they find a mother lode with 200 million tons of gold. Yet, after one year, I want 103 euros back. I don't want gold. I want money! If they cannot give me my rightful money, I will confiscate everything they own. I will offer 2 euros for all their possessions (do they have a better offer somewhere?!). I'll just print 2 extra euros and that's it. Actually it is not even needed to print new money. I get everything. At the end of the year, I get my 100 euro back, I get the gold and mining equipment, and they still keep a debt of 1 euro.\\
A loan can only be paid back if the borrower can somehow produce the \emph{same} (!) commodity that is used in the loan, so that it can give back the loan plus the interest. If gold is lent, and the borrower cannot produce gold, he cannot give back the gold plus interest. The borrower will go bankrupt. If, on the other hand, chickens or sacks of grain are borrowed, these chickens or grain \emph{can} be given back with interest.\\
Banks are the only ones that can produce money, therefore the borrowers will go bankrupt. Full stop.

To say it in another way. If we have a system where interest is charged on debt, no way whatsoever can \emph{all} borrowers pay back the money. Somebody \emph{has} to go bankrupt, unless the game of refinancing goes on forever. This game of state financing can go on forever as long as the economy is growing exponentially. That is, it is growing with constant percentage. The national debt, in terms of a percentage of the gross domestic product (GDP) remains constant, if we continuously refinance and increase the debt, as long as the economy GDP grows steadily too. The moment the economy stagnates, it is game over! Debt will rise quickly. Countries will go bankrupt. (Note that increasing debt is thus the result of a stagnating economy and not the other way around!).

The way the system decides who is going bankrupt, is decided by a feed-back system. The first one that seems to be in trouble has more difficulty refinancing its loans ("You have low credit rating. I fear you will not give me back my books. I want a better risk reward. It is now three books for every book borrowed. Take it or leave it! If you don't like it, you can always decide to give me my books now and we'll call it even").

Thus, some countries will go bankrupt, unless they are allowed to let the debt grow infinitely. If not, sooner or later one of them will go bankrupt. In other words, the \emph{average} interest rate is always zero. One way or another. If $x$\% interest is charged, about $x$\% go bankrupt. To be more precise, $y$\% of the borrowed money is never returned, compensating for the $(100-y$\%$)$ that do return it with $x$\% profit. In a mathematical formula: $(1-y/100) \times (1+x/100) = 1$, or $y = 100x/(100+x)$. This percentage goes bankrupt. For example, if 100\% interest is charged, 50\% goes bankrupt.

To take it to the extreme. If the market is cautious -- full of responsible investors -- and decides to lend money only to 'stable' countries, like Germany, which lately (times are changing indeed) has a very good credit rating from the financial speculators, even these 'stable' countries go bankrupt. That is, the weakest of these stable countries. If only money is borrowed to Germany, Germany goes bankrupt. Apart from the technical mathematical certainty that a country can only have a positive trade balance -- essential in getting a good credit rating -- if another country has a \emph{negative} trade balance (the sum, being a balance, is \emph{always} zero). Germany needs the countries like Greece as much as it despises them.\\
Well, in fact, this is not true. A country does not -- nay, it \emph{cannot} -- go bankrupt for money borrowing. Not if it is an isolated country with its own currency, being also the currency in which the money is borrowed. It can simply print money. That is because the money is their own currency based on their own economy.

\section{Money as currency; The euro; State debt}

Money is the base of banking and commerce and thus of the entire economy. The government is managing this economy one way or another. It thus defines the monetary policy. They set the standards of the base money. They also borrow money from the \emph{same} society to finance the governmental operations. Note the important stress on the word 'same'. The fact that borrowers and lenders are from the same society is essential.\\
Government writes out government bonds to finance itself. We should look at it like this: Technically speaking, since the government is the head of society, it is borrowing its own money! It can thus do so at any amount it sees fit. It can reward the lenders with an interest paid, but that is a cigar of their own boxes. At least for the society as a whole. Not so for the individual elements of society as will be shown later. But, imagine for the sake of simplicity that everybody in society lends the same fraction of its money to the state. If the government has to pay interest on it, say 3\%, it just prints 3\% extra bank notes and nobody is the richer, since it will just cause 3\% inflation. We could only hope that an increase in wealth was achieved, and everybody will be happy. Printing banknotes is a form of taxing, transferring buying power from the society (money users) to the money-printing entities.\\
If not everybody lends money, effectively buying power is also transferred from the non-lenders (who feel more inflation than interest) to the lenders (who feel more interest than inflation), in a zero-sum way (interest $-$ inflation = 0). This is a monetary example of how capital attracts capital, which will be discussed in the next section. But, for the moment imagine that everybody is borrowing an equal fraction of its money. Everybody is thus paying for his/her own financing. It is a complete internal effect. Even if the government were to decide to multiply the money in circulation by a factor of 10, this has no effect whatsoever, apart from a 900\% inflation and a 90\% devaluation of the national currency compared to other currencies in the world. Still, having ten times more of that money, this also does not decrease buying power abroad.\\
Government can do with its own currency what it wants. It should not feel ashamed or incommoded to have a budget deficit of $x$\%, or have a national debt of $y$\%. These numbers can even be astronomical. They are just numbers.\\
As an investor it is the safest thing to do to invest in government bonds. Irrespective of what the country does, at the end of the cycle you get a larger share of the country's wealth than if you hadn't invested your money. Without investing, you only suffer from inflation. Putting your money in a mattress is destruction of capital.

It gets totally different with a pan-continent, multi-country, single currency like the euro. In this case, if one region, say Greece, spends more than it earns, citizens from other countries effectively pay for it, one way or another. If money is being printed, inflation occurs that hurts citizens of other countries. That is why a controlled budget is required for all countries. Everybody should have equal deficit in budget, and an arbitrary norm of 3\% was chosen and a state debt of not more than 90\% (the latter being stupidly 'scientifically' proven, which is too silly for words, but this came in very handy by the governments). As a side-mark, even if the government were to have an eternal 3\% budgetary deficit, the national debt, relative to the gross domestic product (GDP), can remain constant, as long as GDP grows with a constant rate as well (not even necessarily at a rate larger than the deficit).

The problem now lies in the fact that this also necessitates the same interest rate for all countries. Unfortunately -- and the cause for all the problems in Europe -- is that only the first part of the euro was introduced (equal budget rules), but not the second (equal interest rates; 'Eurobonds'). Now the market is left to speculate on countries and interest rates fluctuate widely for different countries.

It was thought that exactly this threatening market speculation would force countries to keep within the agreed budgetary limits. However, speculation is a form of a positive feedback system. If one country sticks out just a teeny weeny bit for some reason whatever, the interest rate for this country will go up a little bit and this causes problems for that country and thus more speculation that that country will not make it and more interest rates and more problems, etc. It is an unstable system. Or metastable at best, waiting for an accident to happen. Well, it happened.

Now there are some countries for which the interest rates have sky-rocketed. And there's no way to do anything about it. They cannot print money, since the others won't let them. In fact, it is the others that even benefit from the situation. A country like Germany wins by the misery of other countries. First, directly by usury, lending money for astronomical interest rates. Second, its own interest rates on borrowing money go down, because the investors see it as a safe haven of their money. If it weren't for the fact that the entire economy of Europe is destroyed, affecting also the 'strong' countries, Germany would do nothing, as its money owners find the situation perfect. And government is representing these money lenders and controls its people -- who couldn't care less -- by using simple rhetoric that "these countries deserve the problems they are in because they did something wrong".

\section{FRB 2. State financing by banks}

Inside a bank, as shown in the FRB section, money can be created at will. Imagine now what happens. Germany needs 100 billion euro to finance its budget. It needs to borrow this money. It goes to a bank (or puts it up for sale in an auction, and a bank wins the auction). The bank adds the number 100 billion to the 'account' of the state. Creating money out of air. The state now can make payments from this account. After one year (assuming that is the maturity of the loan), the state pays back the money, plus 3\% interest (assuming that was the interest rate), 103 billion in total. 100 billion is used to cancel the number on the account of the state. 3 billion go to the reserve. The bank has made 3 billion euro on air. The profit rate (ROI, return on investment) for the bank is infinite; zero invested, 3 billion euro profit.\\
If the fractional reserve ratio (RR) is limited, the ROI is less than infinite, but still astronomical. Image the RR is 10\%. 1 billion 'real' money can then be used to lend out 10 billion virtual money. With 300 million interest paid one year later (3\% of 10 billion), the ROI is 30\% (300 million profit for 1 billion invested). A country like Portugal pays 5\% and the RR is often 20. That implies 100\% profit in a year. Banking, the most beautiful thing in the world. Now you understand why a bank like ING (Internationale Nederlanden Groep) does not mind paying a fine of 50\% for early payback of government support. 5 billion fine for two-year-early payback of a 10 billion loan. Any other company would consider the 33\% LOI (loss on investment) a bad deal. Not so for a bank. Apparently they can compensate somehow, and 33\% ROI is rather common in the financial sector.

Even if a bank uses real money for lending and not air, interest on the money makes that all of the money of society will wind up in the bank. FRB just speeds up the process. And if the borrower cannot pay back the money, assets (another form of capital) of this borrower will be confiscated. Generally speaking, interest on money lending is a way of transferring wealth to the capital, as explained by Marx in his classical work The Capital. It is called condensation of wealth.

\section{Marxism, the surplus of labor}

Marxism is a taboo word in our world. Even while you read it, you will now feel some reluctance and urge to ridicule this text, to which you agreed until now. That is because you, like me (I only read these theories very recently), were brought up in that environment. However, Marx has long ago analyzed the phenomenon that wealth is accumulating. This banking system we have is nothing more and nothing less than an example of the phenomenon Marx mentioned in his works.\\
Marx had one brilliant new idea that all the other economists before him missed. He is therefore as brilliant as Darwin is for biology, or Newton for physics. But just like Darwin and Newton, he does not have the final word.\\
His idea is simple, brilliant in its sheer simplicity, and can be summarized in the following question:
\begin{quote}
"Workers produce things, and consume things. Workers produce more than they consume. Who gets the surplus labor (production minus consumption) of the workers?"
\end{quote}
This is undeniably true. Not a iota of doubt is possible. In his work, The Capital, he argues that in the capitalist society (one of the five ways a society can be organized, that is, one of the five ways the distribution of this surplus can be organized) this surplus goes to the 'capital', which thus accumulates more and more wealth.
Here, for historic reasons, the same wording will be used, 'capitalist' is synonym for 'those that accumulate capital/money', without going into detail if, how, when, where and why Marx' observations on the organization of economy are correct or not. My excuses for the wording, if they make you feel queasy, but they are most adequate for describing things.

Marx' famous 'equation' is
\begin{center}
M -- C\{MoP, LP\} -- P -- C' -- M'
\end{center}
Money (M) is used to buy commodities (C, things that can be bought on the market), namely, means-of-production (MoP, tools for the laborers to work with, factories, basic ingredients, etc) and labor power (LP, workers) to do production (P), producing a new commodity (C') and selling it on the market for more money (M'). This money is reinserted on the left and the game re-commences. It is an eternal cycle, whereby the 'profit', M'-M, is originating from skimming of the labor power LP. (Your boss would simply not hire you, if you do not increase his profit margin M'-M. Either be skimmed or go away).

Some people, mostly hardworking small entrepreneurs, oppose to these ideas. They think that \emph{they} are considered by the Marxists as capitalists and that these commies want to take away their hard-earned money. Nothing could be farther from the truth. This is just the propaganda of the system. The hard working small entrepreneurs are just as much victims as the factory workers people have in mind when they think about labor. Small entrepreneurs do produce something and their labor is being skimmed by the system just as much as that of the factory workers. In most cases, entrepreneurs are nearly slaves as well, and the fruit of their work has to be handed over to the system. In fact, the small entrepreneurs are the poorest class of society. They are only incentivated by some kind of illusion that if they work harder, one day they will also have a lot of capital. Indeed in some extremely rare cases, some of them make it there. Once there, they never have to work hard again. The Bill Gates of society. Once a threshold is overcome, there is not even a need to bother to make good products anymore. The capital starts attracting more capital faster than labor can.

\section{Malthusian catastrophe}

As long as wealth is growing exponentially, it does not matter that some of the surplus labor is skimmed. If the production of the laborers is growing x\% and their wealth grows $y$\% -- even if $y$\% $<$ $x$\%, and the wealth of the capital grows faster, $z$\%, with $z$\% $>$ $x$\% -- everybody is happy. This was our situation until recently. The workers minimally increased their wealth, even if their productivity has increased tremendously. Nearly all increased labor production has been confiscated by the capital, exorbitant bonuses of bank managers are an example. (Managers, by the way, by definition, do not 'produce' anything, but only help skim the production of others; it is 'work', but not 'production'. As long as the skimming [money in] is larger than the cost of their work [money out], they will be hired by the capital. For instance, if they can move the workers into producing more for equal pay. If not, out they go).

If the economy is growing at a steady pace ($x$\%), resulting in an exponential growth $(1+x/100)^n$, effectively today's life can be paid with (promises of) tomorrow's earnings, 'borrowing from the future'. (At a shrinking economy, the opposite occurs, paying tomorrow's life with today's earnings; having nothing to live on today).\\
Let's put that in an equation. The economy of today $E_i$ is defined in terms of growth of economy itself, the difference between today's economy and tomorrow's economy, $E_{i+1}-E_i$,
\begin{equation}
E_i = \alpha(E_{i+1}-E_i)
\end{equation}
with $\alpha$ related to the growth rate,  GR $\equiv (E_{i+1}-E_i)/E_i = 1/\alpha$. In a time-differential equation:
\begin{equation}
E(t) = \alpha \frac{{\rm d}E(t)}{{\rm d}t},
\end{equation}
which has as solution
\begin{equation}
E(t) = E_0 e^{1/\alpha},
\end{equation}
exponential growth.

The problem is that eternal growth of $x$\% is not possible. Our entire society depends on a continuous growth; it is the fiber of our system. When it stops, everything collapses, if the derivative ${\rm d}E(t)/{\rm d}t$ becomes negative, economy itself becomes negative and we start destroying things ($E<0$) instead of producing things. If the growth gets relatively smaller, $E$ itself gets smaller, assuming steady borrowing-from-tomorrow factor $\alpha$ (second equation above). But that is a contradiction; if $E$ gets smaller, the derivative must be negative. The only consistent observation is that if $E$ shrinks, $E$ becomes immediately negative! This is what is called a Malthusian Catastrophe, named after Thomas Malthus, a classic economist.\\
Now we seem to saturate with our production, we no longer have $x$\% growth, but it is closer to 0. The capital, however, has inertia (viz.\ The continuing culture in the financial world of huge bonuses, often justified as "well, that is the market. What can we do?!"). The capital continues to increase their skimming of the surplus labor with the same $z$\%. The laborers, therefore, now have a \emph{decrease} of wealth close to $z$\%. (Note that the capital cannot have a decline, a negative $z$\%, because it would refuse to do something if that something does not make profit).

To show you what I mean. Many things that we took for granted before, free health care for all, early pension, free education, cheap or free transport (no road tolls, etc.) are more and more under discussion, with an argument that they are "becoming unaffordable". This label is utter nonsense, when you think of it, since\\
1) Before, apparently, they were affordable.\\
2) We have increased productivity of our workers.\\
1+2 = 3) Things are becoming more and more affordable. Unless, they are becoming unaffordable for some (the workers) and not for others (the capitalists).\\
It might well be that soon we discover that living is unaffordable.

The new money M' in Marx' equation is used as a starting point in new cycle M $\rightarrow$ M'. The eternal cycle causes condensation of wealth to the capital, away from the labor power. M keeps growing and growing. Anything that does not accumulate capital, M'-M $<$ 0, goes bankrupt. Anything that does not grow fast enough, M'- M $\approx$ 0, is bought by something that does, reconfigured to have M'-M large again. Note that these reconfigurations -- optimizations of skimming (the laborers never profit form the reconfigurations, they are rather being sacked as a result of them) -- are presented by the media as something good, where words as 'increased synergy' are used to defend mergers, etc. It alludes to the sponsors of the messages coming to us. Next time you read the word 'synergy' in these communications, just replace it with 'fleecing'.

Important to note: The capital actually 'refuses' to do something if it does not make profit. If M' is not bigger than M in a step, the step would simply not be done, implying also no LP used and no payment for LP. Ignoring for the moment philanthropists, in capitalistic Utopia capital cannot but grow.

If economy is not growing it is therefore always at the cost of labor! Humans, namely, do not have this option of not doing things, because "better to get 99 cents while living costs 1 euro, i.e., 'loss', than get no cent at all [while living still costs one euro]". Death by slow starvation is chosen before rapid death.

In an exponential growing system, everything is OK; Capital grows and reward on labor as well. When the economy stagnates only the labor power (humans) pays the price. It reaches a point of revolution, when the skimming of LP is so big, that this LP (humans) cannot keep itself alive. Famous is the situation of Marie-Antoinette (representing the capital), wife of King Louis XVI of France, who responded to the outcry of the public (LP) who demanded bread (sic!) by saying "They do not have bread? Let them eat cake!"

A revolution of the labor power is unavoidable in a capitalist system when it reaches saturation, because the unavoidable increment of the capital is paid by the reduction of wealth of the labor power. That is a mathematical certainty.

\section{The big pile. (Condensation of wealth)}

The \emph{modus operandi} for this accumulation of wealth to parts of 'the system' (which, for historic reasons, we call 'capitalists') is banking. The 'capitalists' (defined as those that skim the surplus labor of others) accumulate it through the banking system. That is nearly an empty statement, since wealth = money. That is, money is the means of increasing wealth and thus one represents the other. If capitalists skim surplus labor, it means that they skim surplus money. Money is linked to (only!) banks, and thus, accumulation is in the banks.

As shown above, if interest is charged, borrowers will go bankrupt. This idea can be extended. If interest is charged, all money is accumulated in banks. Or, better to say, a larger and larger fraction of money is accumulated in the banks, and kept in financial institutions. The accumulation of wealth is accumulation of money in and by banks. It can only be interesting to see whom the money belongs to.

By the way, these institutions, the capitalists naturally wanting to part with as little as possible from this money, are often in fiscal paradises. Famous are The Cayman Islands, The Bahamas, The Seychelles, etc. Money-Leaks research found a total of 230.000.000.000.000 dollars parked there. That is, about 30 thousand dollars per inhabitant of this planet. Per year, an extra 1 trillion euro ($10^{12}$, a million millions, 6 times the GDP of Portugal) just from the European Union alone is diverted to these fiscal paradises.

With the accumulated money the physical property is bought. Once again, this is an empty statement. Money represents buying power (to buy more wealth). For instance buying the means-of-production (MoP), such as land, factories, people's houses (which will then be rented to them; more money). Etc.\\
Also, a tiny fraction of the money is squandered. It is what normally draws most attention. Oil sheiks that drive golden cars, bunga-bunga parties etc. That, however, is rather insignificant, this way of re-injecting money into the system. Mostly money is used to increase capital. That is why it is an obvious truth that "When you are rich, you must be extremely stupid to become poor. When you are poor, you must be extremely talented to become rich". When you are rich, just let the capital work for you; it will have the tendency to increase, even if it increases slower than that of your more talented neighbor.

To accelerate the effect of skimming, means of production (MoP, 'capital'), are confiscated from everything -- countries and individual people -- that cannot pay the loan + interest (which is unavoidable, as discussed above). Or bought for a much-below market value price in a way of "Take it or leave it; either give me my money back, which I know there is no way you can, or give me all your possessions and options for confiscation of possessions of future generations as well, i.e., I'll give you new loans (which you will also not be able to pay back, I know, but that way I'll manage to forever take everything you will ever produce in your life and all generations after you. Slaves, obey your masters!)"\\
Although not essential (Marx analyzed it not like this), the banking system accelerates the condensation of wealth. It is the \emph{modus operandi}.

Money is accumulated. With that money capital is bought and then the money is re-confiscated with that newly-bought capital, or by means of new loans, etc. It is a feedback system where all money and capital is condensing on a big pile. Money and capital are synonyms.\\
Note that this pile in not necessarily a set of people. It is just 'the system'. There is no 'class struggle' between rich and poor, where the latter are trying to steal/take-back the money (depending on which side of the alleged theft the person analyzing it is). It is a class struggle of people against 'the system'.

There is only one stable final distribution: all money/capital belonging to one person or institute, one 'entity'. That is what is called a 'singularity' and the only mathematical function that is stable in this case. It is called a delta-function, or Kronecker-delta function: zero everywhere, except in one point, where it is infinite, with the total integral (total money) equal to unity. In this case: all money on one big pile.\\
All other functions are unstable.

Imagine that there are two brothers that wound up with all the money and the rest of the people are destitute and left without anything. These two brothers will then start lending things to each-other. Since they are doing this in the commercial way (having to give back more than borrowed), one of the brothers will confiscate everything from the other.\\
Note: There is only one way out of it, namely that the brother 'feels sorry' for his sibling and gives him things without anything in return, to compensate for the steady unidirectional flow of wealth.

\section{Liberalism}

In a humanistic society, boundary conditions ('laws') are set which are designed to make the lives of human beings optimal. Well, at least that is what they are supposed to do, meaning that not always has this goal successfully been achieved. The laws are made by government. Yet, the skimming of surplus labor by the capital is only overshadowed by the skimming by politicians. An average governor gets a pretty handsome salary during his/her mandate and a pension-for-life afterwards. A salary that is actually decided by fellow governors, and excessive payment is chronic (Note that politicians, like managers, do not produce anything. They work, but don't produce). Not many qualities are needed for the job. Being a citizen in most cases suffices; not even being a convicted criminal takes away accessibility to the job, where it prohibitively does so for 'normal' jobs. Moreover, politicians are often 'auto-invited' (by colleagues) in board-of-directors of companies (the capital), further enabling amassing buying power. This shows that, in most countries, the differences between the capital and the political class are flimsy if not non-existent. As an example, all communist countries, in fact, were pure capitalist implementations, with a distinction that a greater share of the skimming was done by politicians compared to more conventional capitalist societies.\\
The reasons for aversion of the people to the political establishments are therefore obvious. People thus prefer liberal systems over centralized systems, since the latter allude to corruption and unfairness, where liberalism, at least on paper, seems very fair.

One form of a humanistic government is socialism, which has set as its goals the welfare of humans. One can argue if socialism is a good form to achieve a humanistic society. Maybe it is not efficient to reach this goal, whatever 'efficient' may mean and the difficulty in defining that concept.

Another form of government is liberalism. Before we continue, it is remarkable to observe that in practical 'liberal' societies, everything is free and allowed, \emph{except} the creation of banks and doing banking. This is left to a small subset of society. Note also that, by definition, a 'liberal government' is a contradiction in terms. A real liberal government would be called 'anarchy'. 'Liberal' is a name given by politicians to make people think they are free, while in fact it is the most binding and oppressing form of government, as will be shown.

Liberalism, by definition, has set no boundary conditions. A liberal society has at its core the absence of goals. Everything is left free; "Let a Darwinistic survival-of-the-fittest mechanism decide which things are 'best'". Best are, by definition, those things that survive. That means that it might be the case that humans are a nuisance. Inefficient monsters. Does this idea look far-fetched? May it be so that in a liberal society, humans will disappear and only capital (the money and the means of production) will survive in a Darwinistic way? Mathematically it is possible. Let me show you.

\begin{quote}
Intermezzo: Trade unions\\
Trade unions are organizations that represent the humans in this cycle and they are the ways to break the cycle and guarantee minimization of the skimming of laborers. If you are human, you should like trade unions. (If you are a bank manager, you can -- and should -- organize yourself in a bank-managers trade union). If you are capital, you do not like them. (And there are many spokesmen of the capital in the world, paid to propagate this dislike). Capital, however, in itself cannot 'think', it is not human, nor has it a brain, or a way to communicate. It is just a 'concept', an 'idea' of a 'system'. It does not 'like' or 'dislike' anything. (Note the amount of apostrophes). You are not capital, even if you are paid by it. Even if you are paid handsomely by it. Even if you are paid astronomically by it. (In the latter case you are probably just an asocial asshole).
We can thus morally confiscate as much from the capital we wish, without feeling any remorse whatsoever. As long as it does not destroy the game; destroying the game would put human happiness at risk by undermining the incentives for production and reduce the access to consumption.\\
On the other hand, the spokesmen of the capital will always talk about labor cost contention, because that will increase the marginal profit M'-M. Remember this, next time somebody talks in the media. Who is paying their salary?\\
To give an idea how much you are being fleeced, compare your salary to that of difficult-to-skim, strike-prone, trade-union-bastion professions, like train drivers and pilots. The companies still hire them, implying that they still bring a net profit to the companies, in spite of their astronomical salaries. You deserve the same salary.
\end{quote}

Continuing. For the capital, there is no 'special place' for human labor power LP. If the Marxist equation can be replaced by
\begin{center}
M -- C\{MoP\} -- P -- C' -- M'
\end{center}
i.e., without LP, capital would do just that, if that is optimizing M'-M. Mathematically, there is no difference whatsoever between MoP and LP. The only thing a liberal system seeks is optimization. It does not care at all, in no way whatsoever, how this is achieved. The more liberal the better. Less restrictions, more possibilities for optimizing marginal profit M'-M. If it means destruction of the human race, who cares? Collateral damage.\\
To make my point: Would you care if you had to pay (feed) monkeys one-cent peanuts to find you kilo-sized gold nuggets? Do you care if no human LP is involved in your business scheme? I guess you just care about maximizing your skimming of the labor power involved, be they human, animal or mechanic. Who cares?

There is only one problem. Somebody should \emph{consume} the products made (no monkey cares about your gold nuggets). That is why the French economist Jean-Baptiste Say said "Every product creates its own demand". If nobody can pay for the products made (because no LP is paid for the work done), the products cannot be sold, and the cycle stops at the step C'-M', the M' becoming zero (not sold), the profit M'-M reduced to a loss M and the company goes bankrupt.

However, \emph{individual} companies can sell products, as long as there are \emph{other} companies in the world still paying LP somewhere. Companies everywhere in the world thus still have a tendency to robotize their production. Companies exist in the world that are nearly fully robotized. The profit, now effectively skimming of the surplus of MoP-power instead of labor power, fully goes to the capital, since MoP has no way of organizing itself in trade unions and demand more 'payment' (that is, demand payment to start with).

Or, and be careful with this step here -- a step Marx could never have imagined -- what if the MoP start consuming as well? Imagine that a factory robot needs parts. New robot arms, electricity, water, cleaning, etc. Factories will start making these products. There is a market for them. Hail the market!

Now we come to the conclusion that the 'system', when liberalized will optimize the production (it is the only intrinsic goal)
Preindustrial (without tools):
\begin{center}
M -- C\{LP\} -- P -- C' -- M'
\end{center}
Marxian
\begin{center}
M -- C\{MoP, LP\} -- P -- C' -- M'
\end{center}
Post-modern
\begin{center}
M -- C\{MoP\} -- P -- C' -- M'
\end{center}
If the latter is most efficient, in a completely liberalized system, it will be implemented. This means\\
1) No (human) LP will be used in production\\
2) No humans will be paid for work of producing\\
3) No human consumption is possible\\
4) Humans will die from lack of consumption\\
In a Darwinistic way humanity will die to be substituted by something else; we are too inefficient to survive. We are not fit for this planet. We will be substituted by the exact things we created. There is nowhere a rule written "liberalism, with the condition that it favors humans". No, liberalism is liberalism. It favors the fittest.

It went good so far. As long as we had exponential growth, even if the growth rate for MoP was far larger than the growth rate for rewards for LP, also LP was rewarded increasingly. When the exponential growth stops, when the system reaches saturation as it seems to do now, only the strongest survive. That is not necessarily mankind. Mathematically it can be either one or the other, without preference; the Marxian equation is symmetrical. Future will tell. Maybe the MoP (they will also acquire intelligence and reason somewhere probably) will later discuss how they won the race, the same way we, Homo Sapiens, currently talk about "those backward unfit Neanderthals".

To finish this part. Your ideal dream job would be to manage the peanut bank, monopolizing the peanut supply, while the peanut eaters build for you palaces in the Italian Riviera and feed you grapes while you enjoy the scenery. Even if you were one of the few remaining humans. A world in which humans are extinct is not a far-fetched world. It might be the result of a Darwinian selection of the fittest.

\section{Living above your standard, condensation of wealth revisited}

\begin{figure}
 \centering
 \scalebox{0.4}{\includegraphics{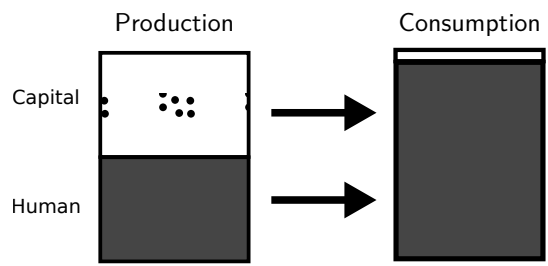}}\\
 \caption{\label{fig:fig1}
Production and consumption of humans and capital
}
\end{figure}

It is often said to countries in trouble that their people were living above their standards. That their consumption is higher than their production. This, in fact, is true ... for everybody on this planet. In financial terms.\\
Look at the image of Fig.\ \ref{fig:fig1}. People (LP), together with machines from the capital (MoP) produce goods that only (mostly) humans consume. Left the production, right the consumption.

\begin{figure}
 \centering
 \scalebox{0.4}{\includegraphics{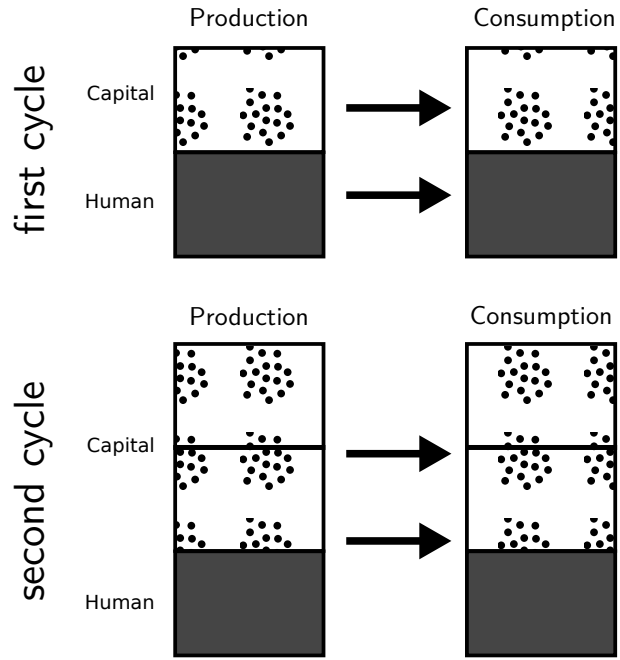}}\\
 \caption{\label{fig:fig2}
Production and consumption of humans and capital, if the capital consumes as humans, but this consumption is used as new starting capital in a new cycle
}
\end{figure}

If everything that is produced is consumed (according to Jean Baptiste Say), it is obvious that humans consume more than they produce. This seems contradictory with the ideas of Marx, but it isn't. Marx said that LP with the \emph{help} of MoP produces, and that this production is fully attributed to LP and is thus skimmed when it consumes less than this production. We can also equally well say that MoP ('capital') is producing with the help of LP, as discussed earlier. Or just say that both are producing and say that each is the right 'owner' of its own production.\\
In the above image, the arrows show the flow of production-consumption. The payment for produced products is an arrow in opposite direction. In this example, humans get 95\% of consumption while they do only 50\% of the production. They thus also only get 50\% of payment. The rest of the consumption is paid by 'borrowing' money somehow, and they live above their standard. The payment goes 50\% to the capital. But, because capital does not consume, this payment is used to increase the capital. Two extreme scenarios:
\begin{itemize}
\item
The money for payment of production is fully in the form of a loan to the humans. Money starts thus accumulating at the capital.
\item
The money for payment is fully used to invest in new capital. In that case, the 'consumption' of capital is 50\%, but after one cycle, a larger part of the production is done by capital. See Figure \ref{fig:fig2}. In the first step, 50\% of the production and consumption is done by capital. In the second cycle it is already 67\%. In the third cycle it is 80\%, then 89\%, etc. In general $2^{n-1}/(2^{n-1}+1)$ at step $n$; capital doubles at every cycle, where humans stay constant. The final situation is that 100\% of production is done by capital. Obviously, sooner or later the system has to switch to the first scenario.
\end{itemize}

In either scenario, the capital accumulates. The basic ingredient is that capital does not need consumption for its survival; any 'consumption' is directly converted into more capital. The system will probably have a mix of the two. After all, capital cannot go on doubling all the time.

So, we see that capital is condensing at the capital. That is because the means of production -- other than human labor -- do not consume, and, therefore, humans do consume more than they produce, and the means of production (machines) do consume less than they produce, with the total in a zero-sum-game way consuming exactly what they produce. The owners of the means of production get the rights to consumption and these rights are constantly increasing. It is a positive-feedback run-away system.

Let's put this in an example to explain it better.
Imagine I make clothespins and so does my neighbor. However, my neighbor has slightly more costs than me, or is slightly less productive for some reason (work accident, or so). He earns just enough to survive. He makes one 'unit' and this barely covers the cost of life, which is also minimally 1 unit. I am slightly more productive, or my cost of living is slightly lower. Therefore, I can save a little 'money'. Let's assume the former, I am more productive. Now, either I make 1.1 units and the surplus 0.1 units I trade for a clothespin machine, or I work a little less on making clothespins and in this spare time -- one hour per day -- I make the machine myself. Let's assume the second scenario, because it is easier reasoning, although they are equivalent. We both make two 'units' of pins, sell them and buy things (two units worth) to survive.
 I however, make as well a machine that makes pins.

After finishing my machine, maybe after ten years, the total production goes up. The demand for our pins stays the same. The markets needs two units of clothespins. It now means that I will get more share of the profit. Imagine my machine makes as much units as a human can, one unit per year. We thus have three units to offer to the market. The price of pins on the market could (and will) drop through the mechanism of supply and demand. In principle down to 67\% of the original price. Not lower, because that would imply that the total price of more pins would be lower than before. 

To make it simple, imagine exactly that happens. The price is 2/3; one unit of pins gives only 2/3 consumption rights. We sell three units and thus get a total of two units of consumption rights. These are distributed over the production units. My neighbor has one third of the production units and thus gets 1/3 share of the consumption rights, a total of 2/3 units. I and my machine get 2/3 share, 4/3 consumption rights. Note that I confiscate -- skim -- the production rights of my 'slave' machine.

Now my neighbor has a problem. He gets 2/3 units of consumption rights, there where one full unit is needed to survive. He did not start working less, or become less productive, or lazy. He simply lost his percentage share of the means of production. And once this starts, there is no stopping it. It in fact accelerates.

There are two scenarios. Either I keep producing pins myself, as shown above, resulting in immediate misery for my neighbor, or I stop working altogether on making pins manually, and we go back to the situation where we make two units of pins, sell them, and each one gets one unit of consumption rights. However, now I have 100\% free time (my machine doing all the work), and I can dedicate it to make a new machine. This takes only one year instead of ten, since I now have 100\% free time, instead of only 10\%. 
In the first situation, I could lend 1/3 of my consumption rights to my neighbor. However -- nothing is for free in this life -- next year I want 10\% profit on my loan. His problems will be bigger next year. Next year I will refinance his loan. Etc. The reader will easily understand that my neighbor will wind up being my feudal possession. I will take everything he owns.
Instead, I could opt for the second path, producing a new machine in my spare time. In that case, next year we will have 4 production units, my neighbor and I as human labor, and two mechanical units. These mechanical units are mine and will claim the consumption rights; together with my own labor, I will now get 75\% of the two consumption rights. 1.5 for me and 0.5 for my neighbor. This path leads to the state where I have 100\% of the consumption rights. Or I can again decide to use part or all of my human labor or machine power to make new machinery. Sooner or later, anyway, my neighbor will have to borrow consumption rights from me.

This is a feedback system. Any small perturbation (we started with 10\% imbalance) results in a saturation in which I will get 100\% of the consumption rights and where I will wind up being the feudal lord of my neighbor. One could argue that this reasoning does not work, because the rest of the world is also increasing productivity and the price of the products offered by them (and the cost of living for me and my neighbor) goes down, as fast as the price of our clothespins go down and we will both easily survive. First of all, we consider here only the local effect, independent of the full market. Technological innovation creates immediate misery for some, a deterioration of life while these people are doing nothing worse. Second, when the rest of the market is behaving in the same way, we remain with an overall effect of condensation of wealth. Capital attracts capital. This is a form of the Matthew Effect, named after the apostle from the bible, transferring money from the poor to the rich. Matthew 25:29, "For onto everyone that hath shall be given, and he shall have abundance, but from him that hath not shall be taken away even that which he hath".

\section{The Pareto-optimal solution}

There are some solutions. ("If you don't give a solution, you are part of the problem").

Most important: Human wealth should be set as the only goal in society and economy. Liberalism is \emph{ruinous} for humans, while it may be optimal for fitter entities.

Nobody is out there to take away the money of others without working for it. In a way of 'revenge' or 'envy', (basically justifying laziness) taking away the hard-work earnings of others. No way. Nobody wants it. Thinking that yours can be the only way a rational person can think. Anybody not 'winning' the game is a 'loser'. Some of us, actually, do not even want to enter the game.\\
Yet -- the big dilemma -- that money-grabbing mentality is essential for the economy. Without it we would be equally doomed. But, I'll show you now that you'll will lose every last penny either way, even without my intervention.

Having said that, the solution is to take away the money. Seeing that the system is not stable and accumulates the capital on a big pile, disconnected from humans, mathematically there are two solutions:\\
1) Put all the capital in the hands of people. If profit is made M'-M, this profit falls to the hands of the people that caused it. This seems fair, and mathematically stable. However, how the wealth is then distributed? That would be the task of politicians, and history has shown that they are a worse pest than capital. Politicians, actually, always wind up representing the capital. No country in the world ever managed to avoid it.\\
2) Let the system be as it is, which is great for giving people incentives to work and develop things, but at the end of the year, redistribute the wealth to follow an ideal curve that optimizes both wealth and increments of wealth.

The latter is an interesting idea. Also since it does not need rigorous restructuring of society, something that would only be possible after a total collapse of civilization. While unavoidable in the system we have, it would be better to act pro-actively and do something \emph{before} it happens.\\
Moreover, since money is air -- or worse, vacuum -- there is actually nothing that is 'taken away'. Money is just a right to consume and can thus be redistributed at will if there is a just cause to do so. In normal cases this euphemistic word 'redistribution' amounts to theft and undermines incentives for work and production and thus causes poverty. Yet, if it can be shown to actually \emph{increase} incentives to work, and thus increase overall wealth, it would need no further justification.

\begin{figure}
 \centering
 \scalebox{0.5}{\includegraphics{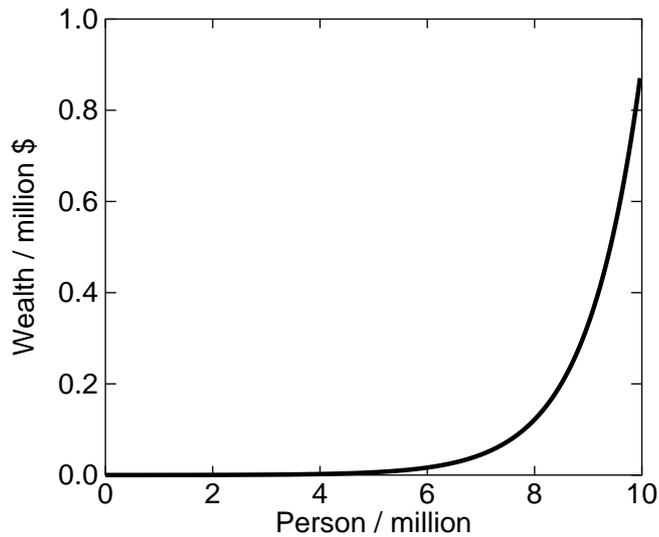}}\\
 \caption{\label{fig:absdistriwealth}
Absolute wealth distribution function
}
\end{figure}

We set out to calculate this idea. However, it turned out to give quite remarkable results. Basically, the optimal distribution is slavery. Let us present them here.\\
Let's look at the distribution of wealth. Figure \ref{fig:absdistriwealth} shows a curve of wealth per person, with the richest conventionally placed at the right and the poor on the left, to result in what is in mathematics called a monotonously-increasing function. This virtual country has 10 million inhabitants and a certain wealth that ranges from nearly nothing to millions, but it can easily be mapped to any country.

As shown in the previous section, the overall wealth increases, but it condenses over time at the right side of the curve. Left unchecked, the curve would become ever-more skew, ending eventually in a straight horizontal line at zero up to the last uttermost right point, where it shoots up to an astronomical value. The integral of the curve (total wealth/capital M) always increases, but it eventually goes to one person.\\
Here it is intrinsically assumed that wealth, actually, is still connected to people and not, as it in fact is, becomes independent of people, becomes 'capital' autonomously by itself. If independent of people, this wealth can anyway be without any form of remorse whatsoever be confiscated and redistributed. Ergo, only the system where all the wealth is owned by people is needed to be studied. This will be done here.

\begin{figure}
 \centering
 \scalebox{0.5}{\includegraphics{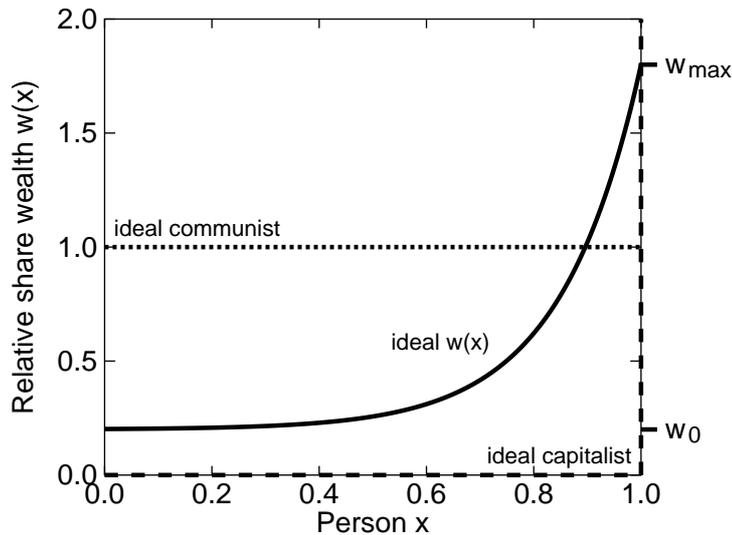}}\\
 \caption{\label{fig:reldistriwealth}
Relative wealth distribution functions: 'ideal communist' (constant distribution, dotted line), 'ideal capitalist' (one person owns all, dashed line) and 'ideal' functions (work-incentive optimized, solid line)
}
\end{figure}

A more interesting figure is the fractional distribution of wealth, with the normalized wealth $w(x)$ plotted as a function of normalized population $x$ (that thus runs from 0 to 1). Once again with the richest plotted on the right. See Figure \ref{fig:reldistriwealth}.\\
Every person $x$ in this figure feels an incentive to work harder, because it wants to overtake his/her right-side neighbor and move to the right on the curve. We can define an incentive $i(x)$ for work for person $x$ as the derivative of the curve, divided by the curve itself (a person will work harder proportional to the \emph{relative} increase in wealth)

\begin{equation}
i(x) = \frac{{\rm d}w(x)/{\rm d}x}{w(x)}.
\end{equation}

To give you an idea. A 'communistic' (in the negative connotation) distribution is that everybody earns equally, that means that $w(x)$ is constant, with the constant being one
\begin{quote}
'ideal' communist: $w(x) = 1$.
\end{quote}
and nobody has an incentive to work, $i(x)$ = 0 for all $x$. However, in a utopic capitalist world, as shown, the distribution is 'all on a big pile'. This is what mathematicians call a delta-function
\begin{quote}
'ideal' capitalist: $w(x) = \delta(x-1)$,
\end{quote}
and once again, the incentive is zero for all people, $i(x) = 0$. If you work, or don't work, you get nothing. Except one person who, working or not, gets everything. Both distributions are also plotted in the figure.

Thus, there is somewhere an 'ideal curve' $w(x)$ that optimizes the sum of incentives $I$ defined as the integral of $i(x)$ over $x$.
\begin{equation}
I = \left. \int_0^1 i(x){\rm d}x = \int_0^1\frac{{\rm d}w(x)/{\rm d}x}{w(x)}{\rm d}x = \int_{x=0}^{x=1}\frac{{\rm d}w(x)}{w(x)} = \ln[w(x)]\right|_{x=0}^{x=1}
\end{equation}
Which function $w$ is that? Boundary conditions are
\begin{itemize}

\item The total wealth is normalized: The integral of $w(x)$ over $x$ from 0 to 1 is unity.
\begin{equation}
\int_0^1 w(x){\rm d}x = 1.
\end{equation}

\item Everybody has a at least a minimal income, defined as the survival minimum. (A concept that actually many societies implement). We can call this $w_0$, defined as a percentage of the total wealth, to make the calculation easy (every year this parameter can be reevaluated, for instance when the total wealth increased, but not the minimum wealth needed to survive). Thus, $w(0)$ = $w_0$.

\end{itemize}

The curve also has an intrinsic parameter $w_{\rm max}$. This represents the scale of the figure, and is the result of the other boundary conditions and therefore not really a parameter as such. The function basically has two parameters, minimal subsistence level $w_0$ and skewness $b$.

As an example, we can try an exponentially-rising function with offset that starts by being forced to pass through the points (0, $w_0$) and (1, $w_{\rm max}$):
\begin{equation}
w(x) = w_0 + (w_{\rm max}-w_0)\frac{e^{bx}-1}{e^b-1}.
\end{equation}
An example of such a function is given in Figure \ref{fig:reldistriwealth}. To analytically determine which function is ideal is very complicated, but it can easily be simulated in a genetic algorithm way. In this, we start with a given distribution and make random mutations to it. If the total incentive for work goes up, we keep that new distribution. If not, we go back to the previous distribution.

\begin{figure}
 \centering
 \scalebox{0.5}{\includegraphics{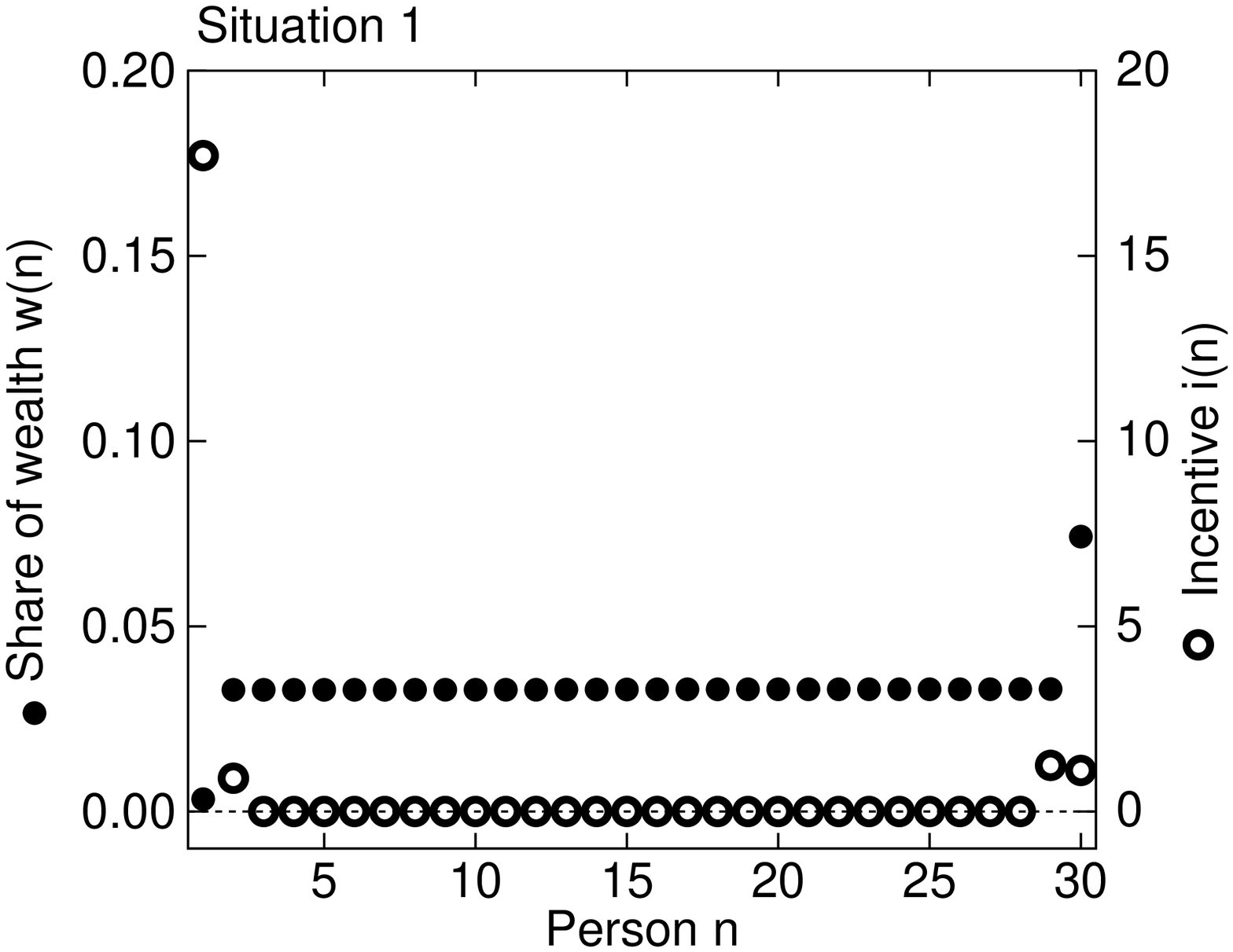}}
 \scalebox{0.5}{\includegraphics{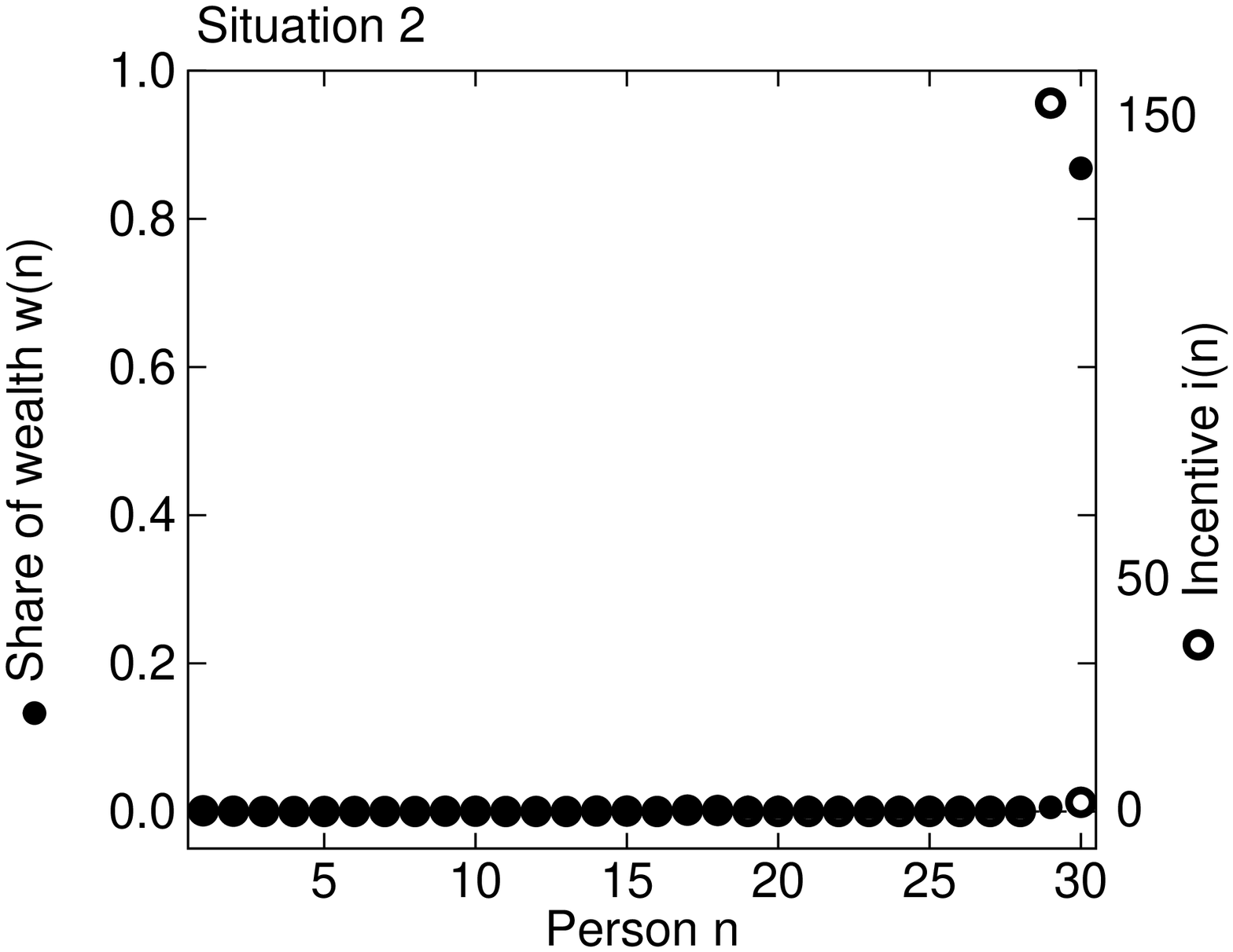}}\\
 \caption{\label{fig:simresult}
Genetic algorithm results for the distribution of wealth ($w$) and incentive to work ($i$) in a liberal system where everybody only has money (wealth) as incentive
}
\end{figure}

The results are shown in Figure \ref{fig:simresult} for a 30-person population, with $w_0$ = 10\% of average ($w_0$ = 1/300 = 0.33\%). Depending on the starting distribution, the system winds up in different optima. If we start with a communistic distribution of Fig.\ \ref{fig:reldistriwealth}, we wind up with a situation in which the distribution stays homogeneous 'everybody equal', with the exception of two people. A 'slave' earns the minimum wages and does nearly all the work, and a 'party official' that does not do much, but gets a large part of the wealth. Everybody else is equally poor (total incentive/production equal to 21), $w$ = 1/30 = 10$w_0$, with most people doing nothing, nor being encouraged to do anything. (See situation 1 of Fig.\ \ref{fig:simresult}). The other situation we find when we start with a random distribution or linear increasing distribution. The final situation is shown in situation 2 of the figure. It is equal to everybody getting minimum wealth, $w_0$, except the 'banker' who gets 90\% (270 times more than minimum), while nobody is doing anything, except, curiously, the penultimate person, which we can call the 'wheedler', for cajoling the banker into giving him money. The total wealth is higher (156), but the average person gets less, $w_0$.

Note that this isn't necessarily an evolution of the distribution of wealth over time (something which was discussed in the section on Marxism). Instead, it is a final, stable, distribution calculated  with an evolutionary ('genetic') algorithm. 

Note 2: This analysis can be made within a country, analyzing the distribution of wealth between people of the same country, as well as between countries.

We thus find that a liberal system, moreover one in which people are motivated by the relative wealth increase they might attain, winds up with most of the wealth accumulated by one person who not necessarily does any work. This is then consistent with the tendency of liberal capitalist societies to have indeed the capital and wealth accumulate in a single point, and consistent with Marx' theories that predict it as well. A singularity of distribution of wealth is what you get in a liberal capitalist society where personal wealth is the only driving force of people. Which is ironic, in a way, because by going only for personal wealth, \emph{nobody} gets any of it, except the big leader. It is a form of Prisoner's Dilemma.

\begin{quote}
Intermezzo: Prisoner's Dilemma\\
A system suffering from Prisoner's Dilemma cannot find the optimal solution because the individual driving forces go against the overall driving force. This is called Prisoner's Dilemma based on the imaginary situation of two prisoners:\\
Imagine two criminals, named Albert (A) and Barbara (B), being caught and put in separate prison cells. The police is trying to get confessions out of them. They know that if none will talk, they will both walk out of there for lack of evidence. So the police makes a proposal to each one: "We'll make it worth your while. If you confess, and your colleague not, we give you 10 thousand euro and your colleague will get 50 years in prison. If you both confess you will each get 20 years in prison". The decision table for these prisoners is like this:

\begin{center}
\begin{tabular}{|c|l|l|}
\hline
Confessions & \multicolumn{1}{c|}{A yes} &  \multicolumn{1}{c|}{A not}\\
\hline
\multirow{2}{*}{B yes} & A gets 20 years,  & A gets 50 years,\\
& B gets 20 years & B gets 10 k\euro\\
\hline
\multirow{2}{*}{B not} & A gets 10 k\euro, & A walks free,\\
& B gets 50 years & B walks free\\
\hline
\end{tabular}
\end{center}

\noindent
As you can see for yourself, the individual option for Albert, independent of what Barbara decides to do, is confessing; moving from right column to left column, it is either reducing his sentence from 50 to 20 years, or instead of walking out of there even getting a fat bonus on top. The same applies to Barbara, moving from bottom row to top row of the table. So, they wind up both confessing and getting 20 years in prison. That while it is obvious that the optimal situation is both not talking and walking out of prison scotfree (with the loot!). Because Albert and Barbara cannot come to an agreement, but both optimize their own personal yield instead, they both get severely punished!
\end{quote}

The Prisoner's Dilemma applies to economy. If people in society cannot come to an agreement, but instead let everybody take decisions to optimize the situation for themselves (as in liberalism), they wind up with a non-optimal situation in which all the wealth is condensed on a single entity. This does not even have to be a person, but the capital itself. Nobody will get anything, beyond the alms granted by the system. In fact, the system will tend to reduce these alms -- the minimum wages, or unemployment benefit, $w_0$ -- and will have all kinds of dogmatic justifications for them, but basically is a strategy of divide-and-conquer, inhibiting people to come to agreements, for instance by breaking the trade unions.\\
An example of a dogmatic reason is "lowering wages will make that more people get hired for work". Lowering wages will make the distortion in the above figures more severe. Nothing more. Moreover, as we have seen, work can be done without human labor. So if it is about competition, men will be cut out of the deal sooner or later. It is not about production. It is about who gets the rights to the consumption of the goods produced. That is also why it is important that people should unite, to come to an agreement where everybody benefits. Up to and including the richest of them all! It is better to have 1\% of 1 million than 100\% of 1 thousand.

Imagine this final situation: All property in the world belongs to the final pan-global bank, with their headquarters in an offshore or fiscal paradise. They do not pay tax. The salaries (even of the bank managers) are minimal. So small that it is indeed not even worth it to call them salary.

\section{Back to banking 1. Houses as the base of economy}

There are still some observations about interesting aspects of banking and the economy that can be made.

The first aspect is the link between banking and houses. In most countries, lending of money is done on basis of property, especially houses. As collateral for the mortgage, often houses are used. If the value of the house increases, more money can be borrowed from the banks and more money can be injected into society. More investments are generally good for a country. It is therefore of prime importance for a country to keep the house prices high.

The way this is done, is by facilitating borrowing of money, for instance by fiscal stimulation. Most countries have a tax break on mortgages. This, while the effect for the house buyers of these tax breaks is absolutely zero. That is because the price of a house is determined on the market by supply and demand. If neither the supply nor the demand is changing, the price will be fixed by 'what people can afford'. Imagine there are 100 houses for sale and 100 buyers. Imagine the price on the market will wind up being 100 k\euro, with a mortgage payment (3\% interest rate) being 3 thousand euro per year, exactly what people can afford. Now imagine that government makes a tax break for buyers stipulating that they get 50\% of the mortgage payment back from the state in a way of fiscal refund. Suddenly, the buyers can afford 6 thousand euro per year and the price on the market of the house will rise to 200 thousand euro. The net effect for the buyer is … zero. Yet, the price of the house has doubled, and this is a very good incentive for the economy. This is the reason why nearly all governments have tax breaks for home owners.

Yet, another way of driving the price of houses up is by reducing the supply. Socialist countries made it a strong point on their agenda that having a home is a human right. They try to build houses for everybody. And this causes the destruction of the economy. Since the supply of houses is so high that the value drops too much, the possibility of investment based on borrowing money with the house as collateral is severely reduced and a collapse of economy is unavoidable. Technically speaking, it is of extreme simplicity to build a house to everybody. Even a villa or a palace. Yet, implementing this idea will imply a recession in economy, since modern economies are based on house prices. It is better to cut off the supply (destroy houses) to help the economy.

\section{Destruction of wealth vs. The Invoculator}

This is a general phenomenon in economy. To starve people and deprive them of things is the way the economy will be booming in financial terms. In fact, 'economy' is based on the \emph{scarcity} of things. As an example, the OPEC (oil producing and exporting countries), in order to drive up the price and economy, has concocted a story of the scarcity of oil, just so to become richer.
This is summarized in the following joke of a conversation of a boy with his mother:
\begin{quote}
- Mother, why is it so cold in the house?\\
- Because there is no coal in the stove\\
- Why is there no coal in the stove?\\
- Because your father doesn't have a job\\
- Why father does not have a job?\\
- Because there is too much coal in the world!
\end{quote}

One can go so far as to say that the lower the wealth, the higher the economy. Imagine a country with a single baker that makes only 10 breads per day. Each bread will cost a fortune. Each bread will contribute millions to the GDP. Now imagine a situation in which everybody has a fancy 'invoculator'\footnote{Name based on the song Multiphasic Invoculator of Freaky Chakra from the album Lowdown Motivator (1995)} at home. The invoculator can, with a single button press, create (invoke) an instance of anything the owner desires, including a new invoculator, but also bread. This will obviously be ruinous for the economy!
A government is normally only worried about the economy and not about wealth. While economy is linked to wealth by being the means to increase it, focusing only on economy is a simplistic and single-sided approach.\\
In fact, many government measures are aimed at \emph{reducing} wealth. As an example, the European Union has recently subsidized the destruction of cars. (For this they used the environment as a justification. Something that is funny in itself, since old but clean petrol cars were thus replaced by new and highly polluting diesel cars). The reason they do this is simple: Wealth is not economy. Economy is \emph{acquiring} wealth. Governments are only aiming at increasing economy. Thus, a measure that destroys wealth and then letting people acquire it again, is perfect.\\
Similarly, a new wave of incentives will aim at replacing combustion-engine cars by electric cars. This is very beneficial for the economy. Once again, the justification is the environment, while, once again, this justification does not make sense. Electricity has to be produced by burning fossil fuels and then transported to the cars and stored there. This is more polluting than directly burning fossil fuels in cars. Apart from that, these electric cars have to be produced and the old ones scrapped. (Probably dumped in some backward country).

In case the 'invoculator' one day indeed is invented, without a shred of doubt, government will take the opportunity to limit its use by law and by taxing schemes. And it is not even so far-fetched. For 'ideas', and any intellectual property in general, the invention already exists, and is called internet (Google, YouTube, etc). Music and films can be copied infinitely without the loss of property to anybody. My own lectures can be replaced 100\% by free and copyable products, see for example the perfect lectures of khanacademy.org, or MIT. But also for physical tangible objects we are getting there. 3D printers already start appearing on the market. For sure, the patent on this invention will be granted eternal rights and its use taxed. The same like the copyright of money creation. It is essential that such rights remain restricted to a selected group of people. Giving everybody these rights will be disastrous for the economy. That while a society where everything is for free forever for everybody is feasible. Actually, \emph{more} music, etc., is made in such a society. We do not need to protect the rights of anybody.

While governments are going about optimizing economy and use destruction of wealth as a tool, society should opt for the opposite path, optimizing wealth and using destruction of economy as a tool. For instance my lectures. I should be fired and my lectures replaced by on-line versions. We would have the same amount of people educated (the same wealth), but with less economy. And I can go to the beach, or study something interesting and useful myself.

In contrast, the only way we can keep an economy running is by destruction of wealth. This is an unavoidable consequence.

\section{Back to banking 2. The stock market}

The next item of banking that is interesting is the stock holders. It is often said that the stock market is the axis-of-evil of a capitalist society. Indeed, the stock owners will get the profit of the capital, and the piling up of money will eventually be at the stock owners. However, it is not so that the stock owners are the evil people that care only about money. It is principally the managers that are the culprits. Mostly bank managers.

\begin{figure}
 \centering
 \scalebox{0.4}{\includegraphics{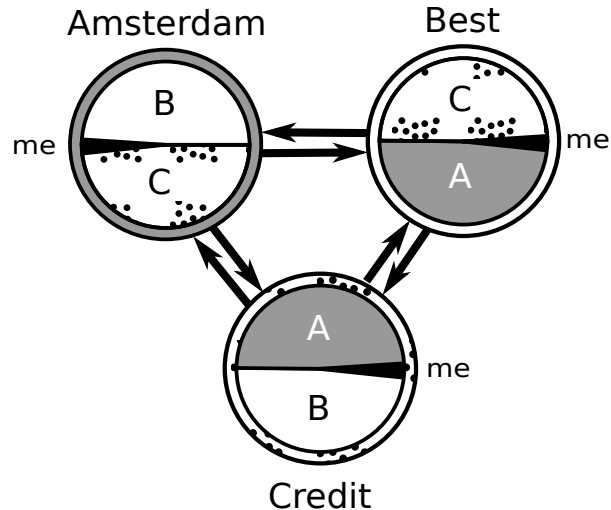}}\\
 \caption{\label{fig:banks}
Three banks with stock placed at each other. The rest of the stock (2\%) is mine. I own all three banks, yet I have nothing to say in them
}
\end{figure}

To give you an example. Imagine I have 2\% of each of the three banks, Amsterdam Bank, Best Bank and Credit Bank. Now imagine that the other 98\% of the stock of each bank is placed at the other two banks. Amsterdam Bank is thus 49\% owner of Best Bank, and 49\% owner of Credit Bank. In turn, Amsterdam Bank is owned for 49\% by Best Bank and for 49\% by Credit Bank. The thing is that I am the full 100\% owner of all three banks. As an example, I own directly 2\% of Amsterdam Bank. But I also own 2\% of two banks that each own 49\% of this bank. And I own 2\% of banks that own 49\% of banks that own 49\% of Amsterdam Bank. This series adds to 100\%. I am the full 100\% owner of Amsterdam Bank. And the same applies to Best Bank and Credit Bank. This is easy to see, since there do not exist other stock owners of the three banks. These banks are fully mine, see Fig.\ \ref{fig:banks}.

However, if I go to a stockholders meeting, I will be outvoted on all subjects. Especially on the subject of financial reward for the manager. If today the 10-million-euro salary of Abel Amberville of Amsterdam Bank is discussed, it will get 98\% of the votes, namely those of Bernard Blacksmith representing Best Bank and Cain Commonnidiyot of Credit Bank. They vote in favor, because next week is the stockholders meeting of their banks. This game only ends when Cain will be angry with Abel.

This structure, placing stock at each other's company is a form of bypassing the stock holders -- the owners -- and allow for plundering of a company.

There is a side effect which is as beneficial as the one above. Often, the general manager's salary is based on a bonus-system; the better a bank performs, the higher the salary of the manager. This high performance can easily be bogus. Imagine the above three banks. The profit it distributed over the shareholders in the form of dividend. Imagine now that each bank makes 2 million profit on normal business operations. Each bank can easily emit 100 million profit in dividend without loss! For example, Amsterdam Bank distributes 100 million: 2 million to me, 49 million to Best Bank and 49 million to Credit Bank. From these two banks it also gets 49 million euro each. Thus, the total flux of money is only 2 million euro. See Table \ref{tab:bankbalance}.

\begin{table}

\caption{\label{tab:bankbalance} Table: Balance sheet of Amsterdam Bank (in millions of euros). Net profit after dividend: zero}
\begin{center}
\begin{tabular}{ll}
\hline\hline
\textbf{Income on banking activity: 2} &\\
\hline
Dividend from Best Bank: 49 &\\
Dividend from Credit Bank: 49 &\\
Total income from dividend: 98 &\\
\textbf{Total profit before dividend: 100} &\\
\hline
& Dividend to Best Bank: 49\\
& Dividend to Credit Bank: 49\\
& Dividend to Peter Stallinga: 2\\
& \textbf{Total dividend paid: 100}\\
\hline\hline
\end{tabular}
\end{center}
\end{table}

Shareholders often use as a rule-of thumb a target share price of 20 times the dividend. This because that implies a 5\% ROI and slightly better than putting the money at a bank (which anyway invests it in that company, gets 5\%, and gives you 3\%). However, the dividend can be highly misleading, as can be seen in the table above. 2 million profit is made, 100 million dividend is paid. Each bank uses this trick. The general managers can present beautiful data and get a fat bonus.

The only thing stopping this game is taxing. What if government decides to put 25\% tax on dividend? Suddenly a bank has to pay 25 million where it made only 2 million real profit. The three banks claimed to have made 300 million profit in total, while they factually only made 6 million; the rest came from passing money around to each other. They have to pay 75 million dividend tax. How will they manage?! That is why government gives banks normally a tax break on dividend (except for small stockholders like me). Governments that like to see high profits, since it also fabricates high GDP and thus guarantees low interest rates on their state loans.

Actually, even without taxing, how will they manage to continue presenting nice data in a year where no profit is made on banking activity?

\section{Conclusions}

We can now answer the two main questions

\noindent
1) Why do we have a crisis?\\
The system, by being liberal, allowed for the condensation of wealth. This went well as long as there was exponential growth and humans also saw their share of the wealth growing. Now, with the saturation, no longer growth of wealth for humans was possible, and actually decline of wealth occurs since the growth of capital has to continue (by definition).\\
Austerity will accelerate this reduction of wealth, and is thus the most-stupid thing one could do. If debt is paid back, money disappears and economy shrinks. The end point will be zero economy, zero money, and a remaining debt.\\
It is not possible to pay back the money borrowed. The money simply does not exist and cannot be printed by the borrowers in a multi-region single-currency economy.

\noindent
2) What will be the outcome?\\
If countries are allowed to go bankrupt, there might be a way that economy recovers. If countries are continuing to be bailed-out, the crisis will continue. It will end in the situation that all countries will have to be bailed-out by each-other, even the strong ones. It is not possible that all countries pay back all the debt, even if it were advisable, without printing money by the borrowing countries.\\
If countries are not allowed to go bankrupt, the 'heritage', the capital of the citizens of countries, now belonging to the people, will be confiscated and will belong to the capital, with its seat in fiscal paradises. The people will then pay for using this heritage which belonged to them not so long time ago, and will actually pay for it with money that will be borrowed. This is a modern form of slavery, where people posses nothing, effectively not even their own labor power, which is pawned for generations to come. We will be back to a feudal system.

On the long term, if we insist on pure liberalism without boundaries, it is possible that human production and consumption disappear from this planet, to be substituted by something that is fitter in a Darwinistic way.

What we need is something that defends the rights and interests of humans and not of the capital, there where all the measures -- all politicians and political lobbies -- defend the rights of the capital. As an example, the 'troika' is a committee representing the interests of the financial system. Nothing more. It is not in Portugal and other countries to help the people. It is in these countries to guarantee that the debt is paid back, whatever it takes.\\
It is obvious that the political structures have no remorse in putting humans under more fiscal stress, since the people are inflexible and cannot flee the tax burden. The capital, on the other hand, is completely flexible and any attempt to increase the fiscal pressure makes that it flees the country. Again, the Prisoner's Dilemma makes that all countries increase tax on people and labor, while reducing the tax on capital and money. The absurd taxing of 2\% of Apple in Ireland is an example. Without a doubt, the employees of Apple in Ireland pay more than 2\% tax.

We could summarize this as saying that the capital has joined forces -- has globalized -- while the labor and the people are still not united in the eternal class struggle. This imbalance makes that the people every time draw the short straw. And every time the straw gets shorter.

\section{Acknowledgements}

This document is the result of many fruitful discussions I had with many people and reading and watching a large quantity of sources. I especially would like to thank the following:

\begin{itemize}

\item
Klaas Bakker, Alan Hollander, Frank Sarlemijn, and Mark de Langen for endless discussions about philosophy (especially dialectics, idealism vs. materialism, and empirical falsification), Marxism and economy in general, banking and politics. Most, actually, could not stand my dialectical techniques anymore and gave up in the later stages of the discussion. For that, my sincere excuses to them.

\item
Paul Krugman for pointing out that Austerity does not work. Never has and never will.

\item
Richard Wolff for the best on-line Marxism lectures one could wish (I still do not understand a thing about the original book of Marx, but the lectures of Wolff are eye-opening and so well given that it should be mandatory viewing material for every student of Economy).

\item
The YouTube movie MoneyAsDebt by paagleTV.

\item
The YouTube series on banking and accounting by khanacademy.

\item
Books: "The Ascent of Money" - Niall Ferguson. "The Black Swan" - Nassim Nicholas Taleb.

\item
Nigel Farage, for being the most straightforward, honest and, besides, entertaining and original politician. He understands the system (and correctly predicted many things), and called names to all the other politicians that don't. Sad to see that the non-enlightened still remain in power.

\item YouTube, Google and Wikipedia in general for developing and making the technology available for free, so that everybody can make this kind of analysis for him/herself. The truth is out there. It is up to us to make sense of the overload of information.

\end{itemize}

%Anti-acknowledgements (political statement):
%To all those politicians that are introducing the Austerity measures: Get off the stage! Get yourself an education and get back when you understand things. The possibility that you actually do know what you are doing, but are destroying our lives on purpose, I will ignore as too scandalous to be true. I will give you the benefit of the doubt if you get out of my face. This work was done in spite of your relentless effort to baffle the scientific system with pseudo scientific, semi-religious and in some points fraudulent scientific steering. You are but spokesmen of 'the system' and do anything to keep it running, even if that visibly creates misery around you. You are not representatives of the people, but representatives of the system that whispers you in the ear what to do at every step.

\end{document}